\documentclass[aps,prl,reprint,superscriptaddress,showpacs,floatfix]{revtex4-2}

\usepackage{amsmath,amssymb,graphicx,bm}
\usepackage[T1]{fontenc}
\usepackage{placeins}
\usepackage{xcolor}
\usepackage{microtype}
\usepackage[plainpages=false,pdfpagelabels,colorlinks=true,linkcolor=red,urlcolor=blue,citecolor=blue,pdftitle={Field-Selected Topological Buffering in a Disordered Skyrmion Crystal},pdfdisplaydoctitle=true]{hyperref}
\graphicspath{}

\begin{document}

\title{Field-Selected Topological Buffering in a Disordered Skyrmion Crystal}

\author{Wen-Yu Su}
\affiliation{Lanzhou Center for Theoretical Physics, Key Laboratory of Quantum Theory and Applications of MoE, Key Laboratory of Theoretical Physics of Gansu Province, and Gansu Provincial Research Center for Basic Disciplines of Quantum Physics, Lanzhou University, Lanzhou, Gansu 730000, China}

\author{Nvsen Ma}
\email{nvsenma@buaa.edu.cn}
\affiliation{School of Physics, Beihang University, Beijing 100191, China}

\author{Chen Cheng}
\email{chengchen@lzu.edu.cn}
\affiliation{Lanzhou Center for Theoretical Physics, Key Laboratory of Quantum Theory and Applications of MoE, Key Laboratory of Theoretical Physics of Gansu Province, and Gansu Provincial Research Center for Basic Disciplines of Quantum Physics, Lanzhou University, Lanzhou, Gansu 730000, China}

\author{Hong-Gang Luo}
\email{luohg@lzu.edu.cn}
\affiliation{Institute of Fundamental Physics and Quantum Technology, Ningbo University, Ningbo, 315211 China}
\affiliation{Lanzhou Center for Theoretical Physics, Key Laboratory of Quantum Theory and Applications of MoE, Key Laboratory of Theoretical Physics of Gansu Province, and Gansu Provincial Research Center for Basic Disciplines of Quantum Physics, Lanzhou University, Lanzhou, Gansu 730000, China}

\date{\today}

\begin{abstract}
Quenched disorder can disrupt crystalline order without immediately destroying the topology of its constituent textures, but the relation between these processes in skyrmion crystals remains unclear. Using large-scale simulations of a triangular-lattice chiral magnet with random Dzyaloshinskii--Moriya interactions, we show that the magnetic field selects between two disordering routes. At high fields, global translational coherence is lost at a weak-disorder scale, while sixfold bond-orientational order survives to a larger disorder strength and the total topological charge remains nearly locked up to a substantially larger scale. The
resulting interval defines a topological buffer containing a
Bragg-glass-like skyrmion regime followed by a skyrmion-glass regime. Finite-size scaling, spatial correlations, defect statistics, and spin autocorrelations support their distinct structural and glassy character. At lower fields, bond-orientational disordering nearly coincides with topological reconstruction, eliminating the skyrmion-glass window and contracting the buffer. These results identify the magnetic field as a control knob for separating crystalline disordering from topological-charge loss and establish topological buffering as a mechanism by which topological textures can remain robust in structurally disordered media.
\end{abstract}

\maketitle
The response of skyrmion crystals to quenched disorder provides  an ideal platform for studying several central problems in condensed-matter physics, including the breakdown of crystalline order and the physics of topological spin textures\cite{Roessler2006,Muehlbauer2009,Yu2010,Heinze2011, NagaosaTokura2013,Tokura2021}. One of the key questions is whether disorder destroys the collective organization of the crystal and the topological character of its constituent skyrmions at the same disorder strength\cite{Hoshino_2018,Chudnovsky_2018,Iroulart_2024,Rosales_2024_dis,Reichhardt_2022,Dey_2020}. If these processes occur at different disorder strengths, crystalline order could disappear while the underlying topological textures remain intact\cite{Reichhardt_2022}. Establishing such a separation is vital for classifying disordered skyrmion phases and has direct implications for skyrmion-based technologies, where structural disorder in real materials is inevitable\cite{NagaosaTokura2013,Fert_2017,Boulle2016,Woo2016}. This connection between fundamental phase behavior and the robustness of information-carrying textures makes the disorder response of skyrmion crystals a problem of broad interest.

Previous studies have established that quenched disorder strongly influences skyrmion crystals, producing lattice distortions and pinning~\cite{Silva_2014,Henderson2022} and modifying depinning and collective transport~\cite{Reichhardt_2015,Reichhardt_2022}. A replica-field theory of collective pinning has further described the low-energy dynamics associated with a magnetic skyrmion glass~\cite{Hoshino_2018,Nishikawa2019,Rosales_2023,Han_2008,Li2019}. Separately, studies of thermal disordering have shown that translational and orientational correlations can evolve differently~\cite{Garanin_2023}. Recent simulations of random-bond disorder also found that periodic spatial order can be disrupted while recognizable chiral textures persist~\cite{Iroulart_2024}. Together, these results reveal the varied responses of skyrmion crystals to disorder, but leave unresolved whether different components of crystalline order break down together or separately from topological reconstruction, and what controls the resulting hierarchy of disorder scales.

\begin{figure*}[!t]
\centering
\includegraphics[width=0.98\textwidth]{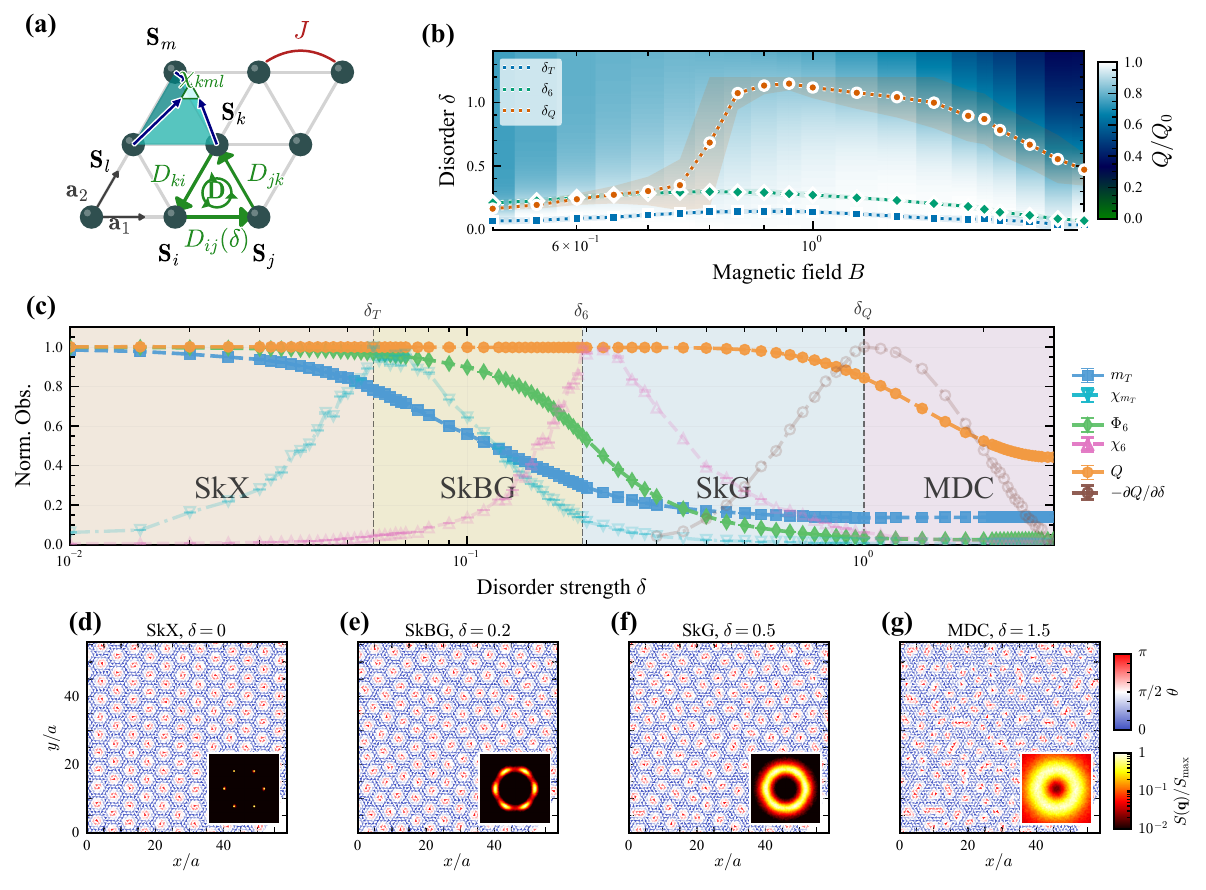}
\caption{\label{fig:phasepath}
\textbf{Triangular-lattice geometry, field-dependent disorder
scales, and texture evolution.}
(a) Triangular lattice with primitive vectors $\mathbf a_1$ and
$\mathbf a_2$, nearest-neighbor exchange $J$, bond-directed DM
interactions $\mathbf D_{ij}$, and the scalar spin chirality associated with an oriented elementary triangle.
(b) Magnetic-field--disorder map obtained on a $90\times90$ lattice. The background color represents the normalized total topological charge $Q/Q_0$. Blue squares, green diamonds, and orange circles denote the characteristic disorder strengths $\delta_T$, $\delta_6$, and $\delta_Q$, respectively.
(c) Normalized translational order parameter $m_T$, sixfold
bond-orientational amplitude $\Phi_6$, and total topological charge
$Q/Q_0$ at $B=1.2$ on a $150\times150$ lattice. The corresponding
translational susceptibility, sixfold susceptibility, and topological response are shown by the lighter curves. Vertical dashed lines mark $\delta_T$, $\delta_6$, and $\delta_Q$. The shaded regions are labeled as the skyrmion crystal (SkX), Bragg-glass-like skyrmion regime (SkBG), skyrmion-glass regime (SkG), and magnetically disordered chiral state(MDC).
(d)--(g) Representative spin configurations at $B=1.2$ on a $60\times 60$ lattice for
$\delta=0$, $0.2$, $0.5$, and $1.5$, respectively. Color denotes the spin polar angle $\theta$, and the insets show the normalized structure factor $S(\mathbf q)/S_{\max}$.
}
\end{figure*}

Here we show that the magnetic field selects between two distinct disordering routes in a two-dimensional triangular-lattice chiral magnet with quenched disorder in the Dzyaloshinskii--Moriya (DM)\cite{Han_2010,Dzyaloshinsky1958,Moriya1960} interactions.
At high fields, global translational coherence is first replaced by roughened, algebraically correlated order, while sixfold bond-orientational order persists to a larger disorder strength and the
total topological charge remains nearly locked up to a still stronger
scale, producing an intermediate skyrmion-glass regime\cite{AragonSanchez2019}. The resulting separation between crystalline disordering and topological reconstruction defines a topological buffer, within which the system can accommodate additional disorder without immediately sacrificing the topological integrity of its skyrmion textures. At lower fields, the loss of bond-orientational order coincides with topological reconstruction into bimeron-rich textures\cite{Silva_2014,Lin2015Meron,Goebel2019Bimeron,Goebel2021Review,Zhang2020Bimeron}, and no skyrmion-glass regime emerges. The magnetic field therefore controls the width of the topological buffer and hence how far topological stability persists beyond the loss of crystalline order.

\textit{Model and observables.---}
We consider classical Heisenberg spins $\mathbf{S}_i$ on a
two-dimensional triangular lattice governed by
\begin{equation}
H=-J\sum_{\langle ij\rangle}\mathbf{S}_i\cdot\mathbf{S}_j
+\sum_{\langle ij\rangle}\mathbf{D}_{ij}\cdot
\left(\mathbf{S}_i\times\mathbf{S}_j\right)
-B\sum_i S_i^z .
\end{equation}
The DM vector is directed along the corresponding nearest-neighbor bond,
$\mathbf{D}_{ij}=(D_0+\delta\eta_{ij})\hat{\mathbf{r}}_{ij}$, where
$\hat{\mathbf{r}}_{ij}=(\mathbf{r}_j-\mathbf{r}_i)/
|\mathbf{r}_j-\mathbf{r}_i|$ and the independent random variables $\eta_{ij}$ are uniformly distributed over $[-1,1]$\cite{GaoRomer2025,Uematsu2021,Rosales_2024_dis,Bilitewski_2017}. The triangular-lattice geometry, exchange and bond-directed DM interactions, and scalar spin chirality\cite{Bilitewski_2017,Rosales_2024_dis,Rosales_2015} associated with an elementary triangle are illustrated in Fig.~\ref{fig:phasepath}(a).

We take $D_0=1$ as the energy unit, set $J=1/3$, and impose periodic boundary conditions. For the fields considered in this Letter, the clean low-temperature state is a triangular skyrmion crystal(SkX)\cite{Binz_2006,Bogdanov_2001,Roessler2006} with translational and sixfold bond-orientational order with a well-defined total topological charge $Q_0$\cite{HalperinNelson1978,NelsonHalperin1979,Strandburg1988}. 
Low-temperature configurations are obtained by simulated annealing down to $T=0.001$ using heat-bath, relaxation, and over-relaxation updates\cite{Metropolis1953,Creutz1987,Kirkpatrick1983,Miyatake1986}.
\begin{figure*}[t]
\centering
\includegraphics[width=0.99\textwidth]{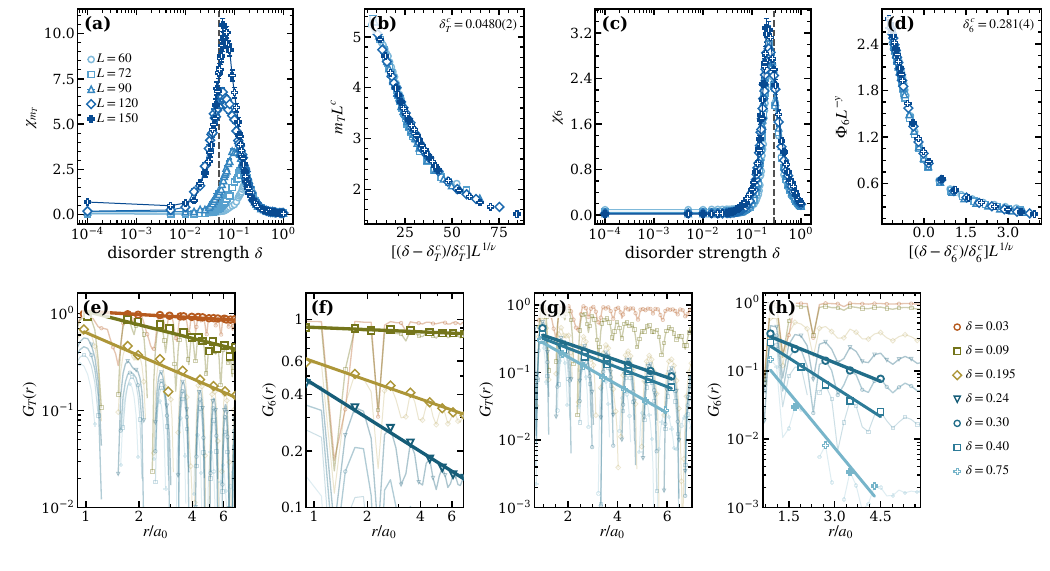}
\caption{\label{fig:braggevidence}\textbf{Finite-size scaling and crystalline correlations at
$B=1.2$}(a),(c) Translational and sixfold susceptibilities,
$\chi_{m_T}$ and $\chi_6$, respectively, as functions of disorder strength $\delta$ for lattice sizes $L=60$, $72$, $90$, $120$, and $150$. (b),(d) Finite-size collapses of the translational order parameter $m_T$ and sixfold bond-orientational amplitude $\Phi_6$, respectively, using the scaled variables indicated on the axes. The optimized collapses give $\delta_T^c=0.0480(2)$ and $\delta_6^c=0.281(4)$. (e),(f) Translational and bond-orientational correlation functions, $G_T(r)$ and $G_6(r)$, displayed on log--log scales for the disorder strengths listed at right.(g),(h) Semilogarithmic representations of $G_T(r)$ and $G_6(r)$, respectively, with exponential fits for the stronger-disorder data. Open symbols identify the data included in the fits, and solid lines show power-law fits in (e),(f) and exponential fits in (g),(h).}
\end{figure*}
To distinguish crystalline disordering from topological reconstruction, we monitor the translational order parameter $m_T$, constructed from the principal reciprocal-lattice components of the skyrmion-density structure factor~\cite{Huang2_020,Nishikawa2019}; the sixfold
bond-orientational amplitude $\Phi_6$
~\cite{HalperinNelson1978,NelsonHalperin1979,Strandburg1988}; and the total topological charge $Q$, evaluated using the
lattice solid-angle construction~\cite{BergLuscher1981,
VanOosterom1983,Rosales_2015}. The characteristic disorder strengths $\delta_T$, $\delta_6$, and $\delta_Q$ are identified from the maxima of the
translational susceptibility $\chi_{m_T}$, the sixfold susceptibility $\chi_6$, and the topological response$-\partial(Q/Q_0)/\partial\delta$\cite{Garanin2023,MendozaCoto2024}, respectively. Full definitions of the observables, simulation protocols, convergence checks, and uncertainty estimates are provided in the Supplemental Material~\cite{SupplementalMaterial}.

\begin{figure}[!tbp]
\centering
\includegraphics[width=1\columnwidth]{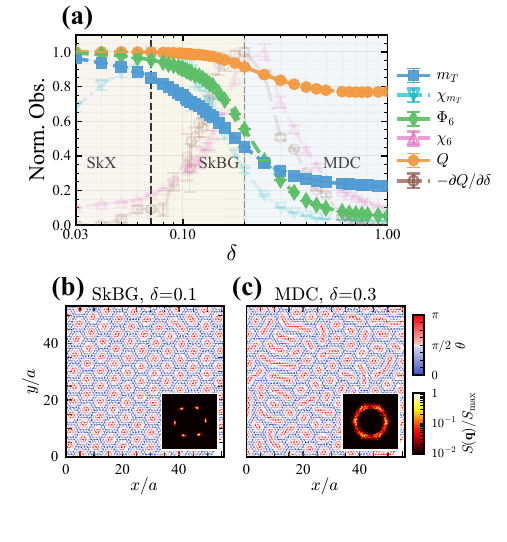}
\caption{\label{fig:lowfield}
\textbf{The texture evolution at lower field.}
(a) Normalized observables along the $B=0.54$ cut on a $150\times150$ lattice.  Translational order is already fragile near weak disorder, while the sixfold susceptibility and topological derivative respond within the same narrow disorder interval, $\delta_6\simeq\delta_Q$, in contrast to the separated high-field route.
(b)--(d) Representative real-space snapshots and corresponding full-configuration structure factors on a $60\times60$ lattice.  Weak disorder mainly disorders positions while preserving recognizable skyrmion cores; stronger disorder elongates and connects the cores, producing a bimeron-rich disordered texture rather than an extended topological buffer.}
\end{figure}

\textit{Field-selected topological buffering.---}
Using these observables, we map the disorder response of the skyrmion
crystal over the $(B,\delta)$ plane. Figure~\ref{fig:phasepath}(b) reveals two
field-selected disordering routes. At intermediate and high fields,
the topological scale $\delta_Q$ lies well above the structural scales $\delta_T$ and $\delta_6$, whereas at lower fields $\delta_6$ approaches $\delta_Q$. 

The high-field route is illustrated by the $B=1.2$ cut in Fig.~\ref{fig:phasepath}(c), which exhibits $\delta_T<\delta_6<\delta_Q$. As the disorder strength increases, translational order changes first near $\delta_T$, followed by bond-orientational order near $\delta_6$, while the normalized topological charge $Q$ remains nearly locked until the disorder approaches $\delta_Q$. We define the interval $\delta_T<\delta<\delta_Q$ as the topological buffer, a finite disorder window in which crystalline organization is progressively degraded before topological reconstruction occurs.

The representative textures and structure factors in Figs.~\ref{fig:phasepath}(d)-\ref{fig:phasepath}(g) illustrate how the system evolves within and beyond the topological buffer. Starting from the skyrmion crystal (SkX), translational order first deteriorates near $\delta_T$,as the Bragg peaks broaden while sixfold bond-orientational order persists, producing a Bragg-glass-like skyrmion regime (SkBG)\cite{GiamarchiPRL1994,Giamarchi1995,Hoshino_2018}.On crossing $\delta_6$,bond-orientational order is also lost and the scattering becomes diffuse, yielding a skyrmion-glass regime (SkG)\cite{AragonSanchez2019}  in which $Q$ remains nearly locked\cite{Hoshino_2018,Chudnovsky_2018}. The disorder strength $\delta_6$ therefore divides the topological buffer into a regime in which bond-orientational order persists after translational order is lost and another regime in which neither component of crystalline order remains on the simulated length scales. Near and beyond $\delta_Q$, the skyrmion cores deform and reconnect as the topological-charge plateau collapses. The resulting decrease in $Q$ indicates that part of the topological charge carried by the original skyrmion textures has been lost, marking the onset of a magnetically disordered chiral(MDC) state \cite{Iroulart_2024}.

\textit{Finite-size scaling and characterization of distinct glassy regimes.---}The separation between $\delta_T$ and $\delta_6$ in Fig.~\ref{fig:phasepath}(c) was identified for a single system size. We therefore examine whether the two disorder scales remain distinct considering the finite-size effect. At $B=1.2$, Fig.~\ref{fig:braggevidence}(a) and \ref{fig:braggevidence}(c) show the translational and sixfold susceptibilities, $\chi_{m_T}$ and $\chi_6$, for several lattice sizes. Their maxima define the size-dependent disorder strengths $\delta_T(L)$ and $\delta_6(L)$. Finite-size collapses\cite{FisherBarber1972,PrivmanFisher1984} of $m_T$ and $\Phi_6$, shown in Fig.~\ref{fig:braggevidence}(b) and \ref{fig:braggevidence}(d), yield the limiting values $\delta_T^c=0.0480(2)$ and $\delta_6^c=0.281(4)$, respectively. Their separation is much larger than the estimated uncertainties, demonstrating that the translational and bond-orientational sectors undergo distinct disorder-driven changes over the accessible system
sizes.

To further characterize the crystalline organization across $\delta_6^c$, we examine the translational and bond-orientational correlation functions, $G_T(r)$ and $G_6(r)$, whose definitions and fitting procedures are provided in the Supplemental Material~\cite{SupplementalMaterial}. In the interval $\delta_T^c<\delta<\delta_6^c$, Figs.~\ref{fig:braggevidence}(e) and \ref{fig:braggevidence}(f) show slow decays consistent with power-law behavior over the accessible distances. Beyond $\delta_T^c$, $G_T(r)$ remains algebraic but becomes progressively weaker, whereas $G_6(r)$ remains comparatively robust up to $\delta_6^c$. This
coexistence of roughened positional order and persistent sixfold bond-orientational correlations supports the identification of this interval as the SkBG regime. For $\delta_6^c<\delta<\delta_Q$, both correlation functions are instead described by exponential decay [Figs.~\ref{fig:braggevidence}(g) and \ref{fig:braggevidence}(h)], indicating finite translational and bond-orientational correlation lengths. Together with the nearly unchanged total topological charge shown in Fig.~\ref{fig:phasepath}(c), this structurally short-ranged regime supports the SkG assignment. The finite-size and correlation analyses therefore resolve two structurally distinct regimes within the topological buffer,  SkBG and SkG, separated by $\delta_6^c$.

Additional evidence for these regime assignments is provided in the Supplemental Material. A Delaunay analysis of the skyrmion-center network reveals dilute non-sixfold sites and predominantly bound five--seven dislocation pairs in the SkBG, followed by a marked increase in defect and dislocation densities near $\delta_6^c$  ~\cite{Rajeswari2015,Gruber2025,LeDoussalGiamarchi2000,
AragonSanchez2019}. Landau--Lifshitz simulations further show that the normalized spin-texture autocorrelation $A(t)$ decays rapidly in a fluid-like reference state but approaches a nonzero long-time plateau in the SkG, providing independent support for its
glassy character~\cite{Udagawa_2016,Gerling_1990,
Bilitewski_2019,Rosales_2023}.

\textit{Lower-field contraction of the topological buffer.---}
We next examine how the disordering route evolves as the magnetic field
is reduced. At $B=0.54$, Fig.~\ref{fig:lowfield}(a) shows that the
translational response occurs at relatively weak disorder, leaving an
interval in which translational order is strongly degraded while
sixfold bond-orientational order and the total topological charge remain
comparatively stable. In contrast to the high-field route, however, the
sixfold susceptibility and topological response peak within the same
narrow disorder interval, giving $\delta_6\simeq\delta_Q$. The
lower-field topological buffer therefore retains a SkBG regime but
contains no separately resolved SkG regime between bond-orientational
disordering and topological-charge loss.

The representative configurations in Figs.~\ref{fig:lowfield}(b) and \ref{fig:lowfield}(c) illustrate this evolution. At $\delta=0.1$, the skyrmion cores remain approximately circular and the scattering retains a pronounced sixfold pattern, consistent with the SkBG regime. With increasing disorder, the sixfold pattern becomes diffuse as the cores deform and reconnect. The accompanying collapse of the topological-charge plateau indicates that the loss of bond-orientational order is accompanied by partial topological-charge loss, taking the system directly from the SkBG into the MDC without an intervening SkG regime.

The magnetic field therefore controls not only the width of the
topological buffer but also its internal structure. At high field, the
buffer contains successive SkBG and SkG regimes, whereas at lower field
it terminates as bond-orientational disordering and topological
reconstruction occur together.

\textit{Conclusion.---}
Our results show that crystalline disordering and topological
reconstruction need not occur at the same disorder strength in a
skyrmion crystal. The separation between these processes defines a topological buffer in which translational and bond-orientational order can be progressively lost while the topological charge remains nearly locked. The magnetic field controls both the extent and the internal structure of this buffer. At high field, distinct SkBG and SkG regimes appear before substantial topological-charge loss, whereas at lower
field the bond-orientational and topological scales approach one another and the SkG regime is no longer resolved. This field tunability suggests a general route for controlling how strongly topological textures can withstand structural disorder.

Several questions follow naturally from this picture. It will be important to determine how the topological buffer evolves at finite temperature, and in the presence of other forms of quenched disorder. The corresponding transport and relaxation responses may provide experimentally accessible signatures of the distinct disordered regimes. More broadly, field-tunable topological buffering may offer a design principle for robust skyrmion memories and reconfigurable information-processing devices intended to operate reliably in structurally imperfect materials. 
\begin{acknowledgments}
This research was supported by the National Natural Science Foundation of China (grant nos.. 12174167, 12474139, 12247101), the Fundamental Research Funds for the Central Universities (Grant No. lzujbky-2025-jdzx07), and the Natural Science Foundation of Gansu Province (No. 25JRRA799).
\end{acknowledgments}

\FloatBarrier

\bibliographystyle{apsrev4-2}
\bibliography{ref,refs_added}

\end{document}